# Design and performance of the LLRF control system for CSNS LINAC

Linyan Rong, Zhencheng Mu, Zhexin Xie, Bo Wang, Wenzhong Zhou, Maliang Wan, Meifei Liu, Jian Li, Xinan Xu

Institute of High Energy Physics, Beijing, China
Spallation Neutron Source Science Center, Dongguan, China

*Abstract*

The China spallation neutron source (CSNS) linac is designed with beam energy of 81MeV and a peak current of 15mA in the first phase. The RF power system for the 81 MeV linac requires 8 units of RF power source, each unit has one independent digital low level RF (LLRF) control system which is used to stabilize the amplitude and phase of the RF accelerating field along the linac, and to minimize beam loss. During beam commissioning, all of 8 on-line units of LLRF control system were operating stably and reliable, the control precision of accelerating fields met requirement of beam commissioning. The amplitude and phase variations of the linac fields are less than ±0.2% and ±0.2° without beam loading, or ±0.5% and ±0.5° with 10mA beam loading, much better than the design requirements of ±1% in amplitude and ±1° in phase.

## INTRODUCTION

The accelerators of China spallation neutron source (CSNS) mainly consist of an H- linac which accelerates the H- beam energy to 81 MeV and a rapid cycling synchrotron (RCS) accelerator which accumulates the proton beam to a high current pulse and then accelerates it to 1.6 GeV. The RF power systems of CSNS 81MeV linac operate at RF frequency 324MHz, repetition rate 25Hz, RF pulse width 650 μs, and duty cycle 1.625% [1]. The RF power system in linac is mainly composed of five klystrons power for RFQ and four DTLs, three solid state amplifiers power for two bunchers and one debuncher. Each unit has one independent digital low level RF (LLRF) control system. The overview of the RF system configuration is showed in Figure 1. The design and performance of the LLRF control system for CSNS linac will be described in this paper.

## LLRF SYSTEM

The CSNS linac LLRF control system is designed to achieve required acceleration voltage amplitude and phase regulation of ±1% and ±1° respectively. The resonant state of the cavity and the beam loading effect should also be taken care of by the low level RF LLRF system, to ensure the stability of the RF system. Each LLRF system is composed of five modules including analog module (AM) and clock distribution module (CDM), LLRF control Module (DCM), high power protection module (HPM), Timing and RF interlock module, the Schematic block diagram of CSNS LLRF is showed in Figure. 2.

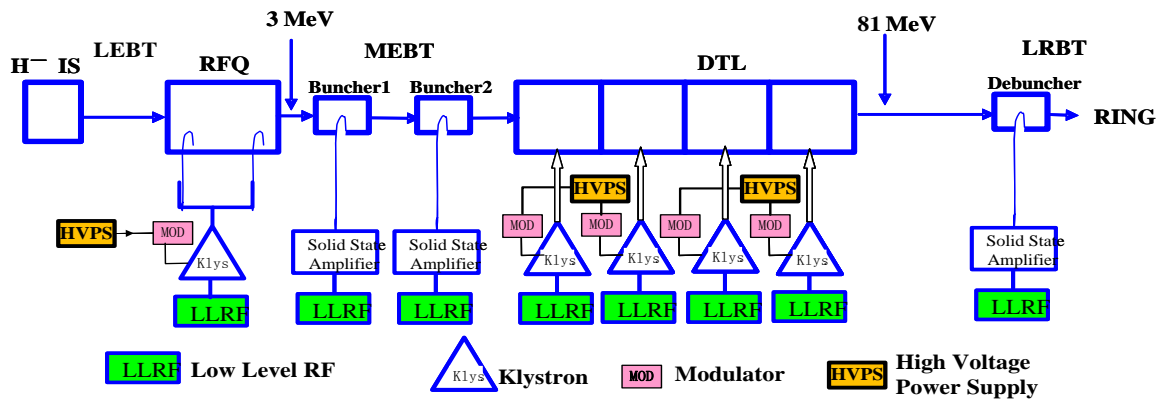

Figure 1: The overview of the CSNS linac RF system configuration

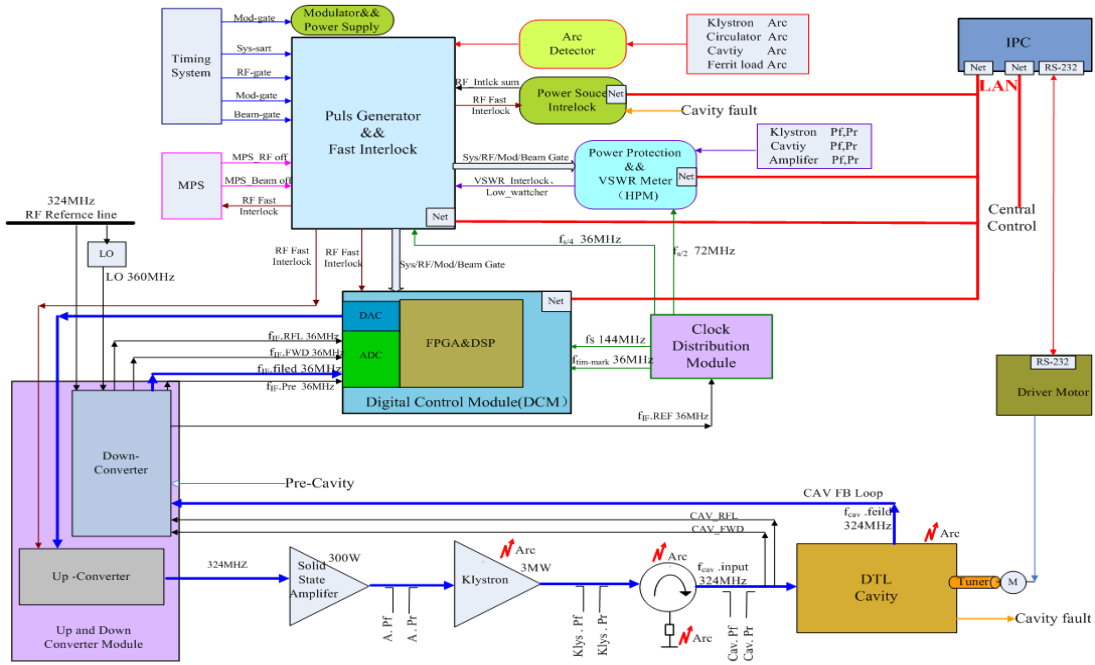

Figure 2 Schematic block diagram of CSNS LLRF

*Analog Module*

The analog module consists of four down conversion channels and two up conversion channels, and the clock generation unit. The clocks distribution module generates three clocks: 144MHz, 36MHz and 72MHz. The 144MHz and 36MHz clocks are used as the 4 channel ADC sampling clock of the DCM, and 72MHz clock is used as the 8 channels ADC sampling clock of the HPM. To avoid it instability, all analog hardware are placed into a temperature stabilizing crate, and the control temperature is maintained at 30 +/- 0.1 degrees Celsius. Semiconductor thermostat is applied in the temperature stabilizing crate, and the regulation is provided by automatic PID bilateral control system with insulation surrounding analog module and RF cables [2].

*RF Reference*

Thermostatically controlled RF reference line of CSNS LINAC will be installed in auxiliary tunnel closed to accelerator cavity in tunnel. 324MHz RF is chosen as the frequency of reference which is the same as the cavity RF. We use R&S SMA 100A signal generator as the MO, MO signal was amplified through low noise amplifier unit, and RF signal was distributed to RF station, Timing system and Beam instrumentations. Reference RF signal and cavity field signal are sent from auxiliary tunnel to LLRF controller through a matched cable pair along the same-length path.

*The Digital Field Control System*

FPGA-based digital LLRF control Module (DCM) is the heart of the LLRF control system. It includes 4 channel ADCs, 4 channel DACs, one Altera straix II FPGA and two TI C6000 DSPs, ethernet interface, the digital signal processing board card as shown in Fig. 3 [3]. The primary control logic is implemented in the high-speed FPGA, the DSPs is mainly responsible for the communication with IPC through Ethernet and some coefficient calculations. This structure can reduce the logic elements usage and simplify the floating point arithmetic, also make the Ethernet communication easy to realize.

In order to increase the speed of digital signal processing, the LO frequency of the LLRF system was designed to be 360MHz, the IF frequency is 36MHz, the ADC sample clock frequency and FPGA signal processing speed can be set to 144MHz, the key controller with all-digital implementation has successfully achieved lower system latency and wider control bandwidth.

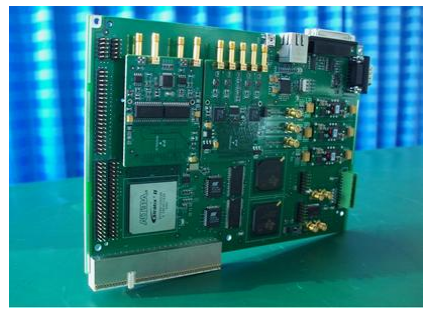

Figure 3: Digital signal processing board card

*Cavity tune loop*

Two methods are adopted to judge whether the RF cavity is close to the resonance state or not in CSNS LLRF control system. The first method is to detect the phase difference between the cavity input and output signals and to see whether it's the same as that for resonance state. The second method is to detect the phase curve of the RF cavity during the field decay and to see whether the phase is a constant.

Since the beam transmission of RFQ accelerator is very sensitive to the field profile, controlling cooling water temperature was adopted to tune the RFQ cavity resonance [4]. There are two working modes, the phase mode and the temperature mode.

Servo motor was adopted to driver the mechanical tuner for DTL cavity, each DTL cavity has two mechanical tuner to maintain the cavity resonant. Displacement sensor sends tunner displacement information to IPC using TCP/IP protocol. The detuning frequency of DTLs is less than 2k during operation.

*The High Power Protection and monitoring*

RF power detection and VSWR protection is implemented by Power Protection Module (PM), Block diagram of PM is shown in Figure 4. The PM module can monitor up to eight RF channels. RF signals are detected by logarithmic detector, and then are sampled by 14bit-ADCs. The fault detection logic is implemented in a single FPGA. Industry PC provides monitoring for eight channels RF power and four channels VSWR.

There are two modes for RF block. The first mode is auto-recovered RF block, if detecting VSWR is over defined threshold, the RF drive and beam shall be inhibited within 1μs of the detection and shall be re-enabled in the next pulse. The second mode is permanent RF block, if detecting VSWR protection take place N times in a second, The RF drive and beam shall be permanently blocked. The response time from detecting VSWR over-limit until the RF drive is inhibited less than 460nS.

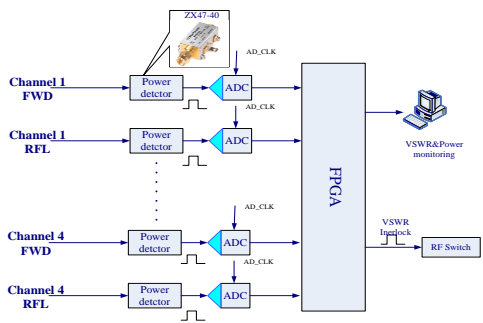

Figure4: Block diagram of PM

## PERFORMANCE

*Stabilization by FB control*

During beam operation，all of 8 on-line units of LLRF control system were reliable and operating stably, the control precision of accelerating fields met requirement of beam commissioning. The amplitude and phase variations of the LINAC fields are less than ±0.2% and ±0.2° without beam loading, or ±0.5% and ±0.5° with 10mA beam loading, much better than the design requirements of ±1% in amplitude and ±1° in phase. FB control result of DTL2 accelerating fields is shown in Figure 5。

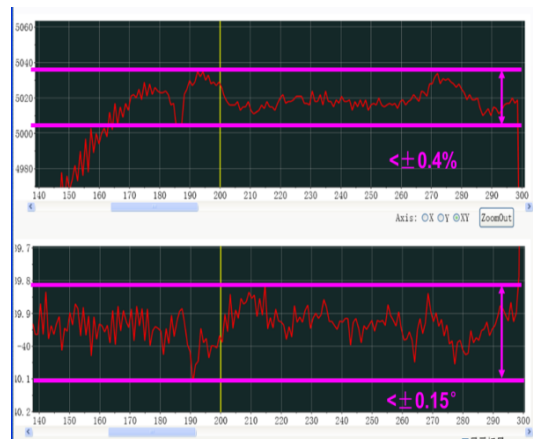

Figure 5: The amplitude and phase curve of DTL2 accelerating fields with beam loading@10mA beam current and 100us beam pulse

*Long Term Stability*

Previous cavity field is used to recover the working point if something changed in the previous RF system. A trend of the amplitude and phase of the DTL4 cavity field under the FB control was measured by Debuncher LLRF during beam operation. The result is shown in Figure 6.The amplitude variation is ±0.6% and the phase variation is ±0.5 degrees during 24h.

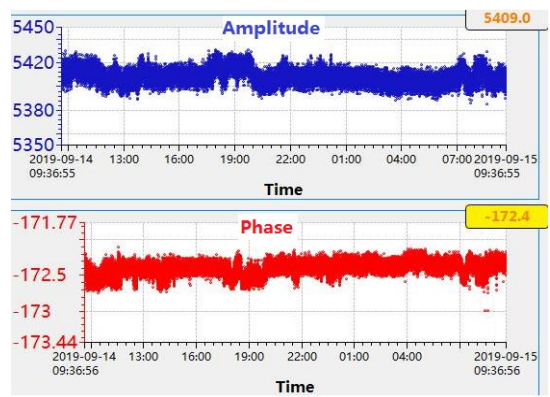

Figure 6: The amplitude and phase trend of the DTL4 cavity field during 24h.

## SUMMARY

LLRF control system began to apply in the operation of the CSNS Linac accelerator at 2015, the control precision of accelerating fields met requirement of beam commissioning. During the several years, LLRF control system had appeared some faults, such as the contact problems of the power plugs, program bug and thermostat chamber faults et al, most of problems have been solved and improved. Also, many important progresses have been achieved in the LLRF control system for a more-convenient operation and a higher stability performance, such as automatic frequency matching for cavity warm-up, the status recovered automatically.

## REFERENCES

[1] LiJian, XuXin'an, MuZhencheng, et al. Linac RF power sources development for China Spallation Neutron Source. High Power Laser and Particle Beams, 2016, 28(8)

[2] Linyan Rong, Yuan Yao, Xinan Xu, et al. The improvements of LLRF control system for CSNS LINAC. LLRF Workshop 2013.

[3] Zhencheng Mu†, Jian Li, Xin An Xu,et al. Overview of the CSNS LINAC LLRF and operational experiences during beam commissioning. Proceedings of HB2016, Malmö, Sweden.

[4] XIN Wenqu, OUYANG Huafu, XU Taoguang. Resonance control cooling system for 973 RFQ at IHEP. Nuclear Physics Review, 2013, 30(2).